\documentclass[onecolumn,showpacs,preprintnumbers,amsmath,amssymb]{revtex4}
\usepackage{graphicx}
\usepackage{dcolumn}
\usepackage{bm}
\usepackage{epsfig}
\usepackage{subfigure}

\newcommand{\bi}{\begin{itemize}}
\newcommand{\ei}{\end{itemize}}
\newcommand{\be}{\begin{eqnarray}}
\newcommand{\ee}{\end{eqnarray}}
\newcommand{\beq}{\begin{equation}}
\newcommand{\eeq}{\end{equation}}
\newcommand{\bbmatrix}{\left( \begin{array}}
\newcommand{\eematrix}{\end{array} \right)}

\begin{document}

\title{An efficient variational approach to the impurity problem and its application to the dynamical mean field theory}

\author{Chungwei Lin and Alexander A. Demkov\footnote{E-mail: demkov@physics.utexas.edu}}
\date{\today}
\affiliation{Department of Physics, University of Texas at Austin}
%538W 120th St NY, NY 10027}

\begin{abstract}
Within the framework of  exact diagonalization (ED), we compute the ground state of Anderson impurity problem 
using the variational approach  based on the configuration interaction (CI) expansion. 
%This approach is based on self-consistently determining the single-particle orbital basis. 
We demonstrate that an accurate ground state can be obtained by
iteratively diagonalizing a matrix with the dimension that is less than 10$\%$  of the full Hamiltonian. 
The efficiency of the CI expansion for different problems is analyzed. 
By way of example, we apply this method to the single-site dynamical mean field theory using ED as the impurity solver. 
Specifically, to demonstrate the usefulness of this approach,
we solve the attractive Hubbard model in the grand-canonical ensemble, where the s-wave superconducting solution is explicitly obtained.
\end{abstract}

\maketitle

%%%%%%%%%%%%%%%%%%%%%%%%%%%%%%%%%%%%%%%%%%%%%%%%%%%%%%%%%%%%%%%%%%%%%%%%%%%%
\section{Introduction}

%A system with particle-particle interaction is generally a very difficult problem to solve.
%For materials with spatially localized orbitals near the Fermi energy, such as transition metal oxides , 
%
The electron-electron interactions are of fundamental importance in wide classes of systems 
and account for many fascinating phenomena, such as the metal insulator transition \cite{RevModPhys.70.1039}, 
superconductivity \cite{RevModPhys.66.763}, and magnetic phase \cite{RevModPhys.79.1015,PhysRevLett.68.851}.
In principle, for a system whose orbitals near the Fermi energy are spatially localized, such as the transition metal oxides
with occupied $d$ orbitals \cite{Hubbard_63}, the electron-electron interaction is important and cannot be neglected.
When taking the inter-particle interaction into account, 
the problem becomes ``strongly correlated'' and is very difficult to solve, because an exponentially large number of states 
couples to one another and a direct determination of the eigenstates becomes computationally impossible.
%whose dimension grows exponentially as the system size increases, couple to one another.
There are several well-developed approaches to the strongly correlated problem. Almost all 
of them, in one way or another, involve identifying a reduced space of the workable dimension 
that includes (or at least overlaps strongly with) the true ground state.
For example, the Hartree-Fock approximation and density functional theory \cite{PhysRev.136.B864,PhysRev.140.A1133} 
map the correlated problem onto a non-interacting one, with the dimension depending linearly on the system size. 
The renormalization approach keeps only low-excited states 
\cite{RevModPhys.47.773,RevModPhys.80.395} in momentum space, or only the most relevant states
\cite{PhysRevB.48.10345,RevModPhys.77.259} in real space. 
Using a variational wave function can greatly reduce the dimension, but this approach is system specific and requires insight into 
the physics of the problem. One of the most successful cases is Laughlin's variational wave function
for the two-dimensional electron gas under a strong magnetic field \cite{PhysRevLett.50.1395}.
If only the ground state properties of the system are of interest, the
recently proposed density matrix  embedding theory requires solving only a two-site problem \cite{PhysRevLett.109.186404}.

The dynamical mean field theory (DMFT) \cite{RevModPhys.68.13, RevModPhys.77.1027, Kotliar_04, RevModPhys.78.865, RevModPhys.83.349}
has emerged as a powerful tool to study the strongly correlated problem.
The key step of DMFT is to map the lattice problem onto a ``simpler'' impurity model \cite{PhysRev.124.41}, 
where a finite set of interacting “impurity” orbitals is coupled to a large number of noninteracting “bath” orbitals.
We will demonstrate in Section III that the impurity problem is simpler in the sense that it requires a much
smaller number of determinantal states to capture the ground state than what is needed in the lattice model.
An accurate numerical solver for correlated impurity models is crucial for DMFT calculations,
and several exact and numerically tractable approaches have been developed. 
The frequently used solvers include the continuous-time quantum Monte Carlo \cite{RevModPhys.83.349},
renormalization \cite{RevModPhys.80.395, PhysRevLett.93.246403},
and exact diagonalization (ED) \cite{PhysRevLett.72.1545,Liebsch_12}. 

Recently, Zgid $et$ $al$. have applied the truncated configuration interaction (CI) expansion, 
a technique developed in quantum chemistry \cite{Helgaker,Sherrill}, to the Anderson impurity model \cite{PhysRevB.28.4315}
and DMFT \cite{Zgid_2011, PhysRevB.86.165128}.
The CI solver can be viewed as an approximation to the full ED,  
as it finds the ground state in the truncated Hilbert space (HS) based on energy considerations.
Zgid $et$ $al$. demonstrate that only a small number of determinantal states is needed to obtain the accurate
ground state of the Anderson impurity model. This finding makes the CI solver a strong candidate for studying 
more complicated problems such as  multi-orbital models, cluster DMFT \cite{PhysRevB.86.165128}, or even 
the non-equilibrium problem \cite{PhysRevLett.110.086403}.
Very recently, the CI method has been applied to the density matrix renormalization group method \cite{Ma_13}. 
%quantum dots \cite{RevModPhys.79.1217}

The CI algorithm and its application to DMFT have been discussed in detail in Ref. \cite{Zgid_2011, PhysRevB.86.165128}.
In this paper, we further analyze the efficiency of the CI method based on the variational principle.
In particular, we formulate an efficient and general procedure, based on the
successive orbital transformation, %, based on self-consistently determining the
%orbital basis using certain class of CI truncation scheme, to obtain the accurate ground state, and 
to accurately express the ground state in a small number of determinantal states,
from which the Green's function can be calculated with minimum computational resources. 
The same procedure determines whether the CI truncation is valid or not. 

The rest of the paper is organized as follows. In Section II we briefly introduce the CI method, and
describe several CI truncation schemes. 
We define the correlation entropy to quantify the degree of correlation.
In Section III we solve an impurity model and an 8-site Hubbard model to demonstrate 
under which circumstances the CI solver can be efficient.
A general strategy to compute the ground state and Green's function is formulated.
In Section IV  we solve the DMFT equation for the Hubbard model with attraction in the grand-canonical ensemble to describe 
the superconducting state.
%In Section V a conclusion is given. 
In the appendices we provide some key steps of successive orbital transformation, and 
details of  solving the impurity problem in the grand-canonical ensemble.

\section{Configuration interaction expansion}

\subsection{Overview and basis classification}
Using the ED to compute the ground state of a given Hamiltonian requires two basic ingredients: 
first, all basis vectors of the HS (or its invariant subspace) have to be labeled, and
second, the matrix elements between any two basis vectors have to be computed.
Once these are available, the ground state can be computed by directly diagonalizing the
Hamiltonian matrix, or by a power method such as Lanczos \cite{RevModPhys.66.763} or 
Davidson algorithms \cite{Davidson197587}. The CI solver \cite{Helgaker,Sherrill,PhysRevB.86.165128} is a 
non-perturbative approximation to solve a many-body problem based on the ED formalism.  
In essence, it is a truncation of HS based on the energy considerations.

\begin{figure}[http]
\epsfig{file=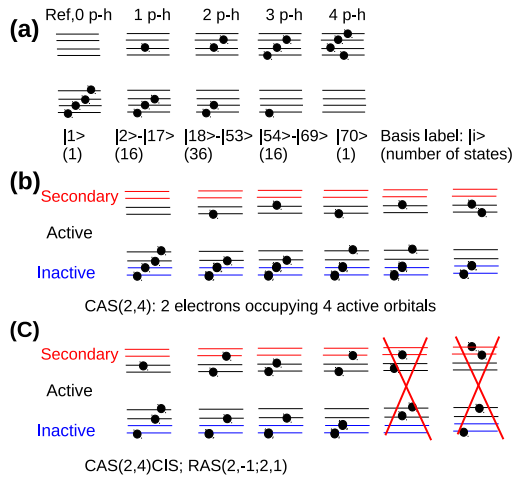, width = 0.4\textwidth}
\caption{(Color online) (a) Basis states for 4 fermions occupying 8 spin-orbitals. Picking up one determinantal state 
as the reference, all basis states can be classified by the number of particle-hole pairs with
respect to the reference state. 
(b) Six determinantal states constructed from CAS(2,4).
Two secondary and active orbitals are always empty and filled respectively.
(c) Some allowed and truncated states in CAS(2,4)CIS or equivalently RAS(2,-1;2,1).
The inactive orbitals can have at most one hole, while secondary orbitals can at most have one particle,
therefore the last two states are not included.
}
\label{fig:StateLabel}
\end{figure}

The many-body fermionic HS is constructed from the antisymmetrized
single-particle orbitals (determinantal states), which are concisely expressed in the second
quantization notation (Fock states). For $n_e$ fermions occupying $n_o$ spin-orbitals, the HS dimension $D_{H}$ is $C^{n_o}_{n_e}
\equiv \frac{n_o!}{n_e! (n_o-n_e)!}$.
As an example, we consider 4 fermions occupying 8 spin-orbitals, with each orbital specified by 
a creation operator $d_j^{\dagger}$ [$j$=1,2,...,$n_o$ ($n_o=8$)]. 
The 4-particle Hilbert space contains $D_{H} = C^8_4=70$ states, with each basis state labeled by
$|i\rangle$ [$e.g.$, $|i\rangle = d_1^{\dagger}d_2^{\dagger}d_3^{\dagger}d_4^{\dagger} |0\rangle$] with $i=1,2,...,70$.
In principle, the ground state is
\beq
|GS\rangle = \sum_{i=1}^{D_H (=70)} c_i |i \rangle.
\label{eqn:GS_general}
\eeq
%This example will be used in this section.
To avoid the confusion in terminology, throughout this paper the ``vectors'' or ``states'' are reserved for the {\em many-body}
wave functions; while ``orbitals'' are reserved for the single-particle wave functions (specified by $\{d_i^{\dagger}\}$)
that are used to build the many-body states. 

By choosing one determinantal state as a reference, all basis states can be classified by
the number of particle-hole (p-h) pairs with respect to the reference state \cite{p-h}. In our example shown in
Fig.~\ref{fig:StateLabel} (a), there are 16 [$(C^4_1)^2$] basis states with one and three p-h pairs, 
36 [$(C^4_2)^2$] with two p-h pairs, and 1 [$(C^4_0)^2$] with zero and four p-h pairs.
The truncated CI expansion only keeps the basis states  with a small number of p-h pairs.
%and it is useful when states of large number of p-h pairs contribute. 
When keeping only the reference state, the CI method is equivalent to the Hartree-Fock approximation.
Keeping states up to 1, 2, 3, and 4 p-h pairs are referred to as CIS, CISD, CISDT, and CISDTQ expansions (S: single, D: double,
T: triple, Q: quadruple), and the corresponding truncated spaces are referred to as the CI subspaces \cite{Helgaker,Sherrill}.
The non-reference states (states with non-zero p-h pairs) are sometimes referred to as ``excitation levels'' in the 
Hartree-Fock sense, but they are not actually the excitations of the Hamiltonian and we avoid using this term here.

The efficiency of the CI truncation %expansion (i.e. to keep minimum number of states)
depends crucially on the choice of the single-particle orbitals. 
As a trivial example to illustrate its importance, 
we note that the many-body ground state of a non-interacting Hamiltonian can be described by a single determinantal state --
if choosing the orbitals diagonalizing the single-particle Hamiltonian, 
all non-reference states can be discarded; otherwise more determinantal states are needed.
%For the rest of this section, we will describe how to choose the orbital basis and 
%eventually to give a criterion determining if CI truncation is valid. 

\subsection{Multi-reference configurations, orbital classification and truncation schemes}

When several determinantal states simultaneously have a significant weight in the ground state, the CI expansion from
a single reference may become inadequate, and a generalization to multi-reference configurations is needed.
To deal with this situation, the {\em orbitals} are classified according to their occupation: 
they are (1) inactive (always occupied); (2) secondary (always empty); or (3) active (partially filled).
Building the CI space based on these three classes of orbitals is referred 
to as a ``complete active space'' (CAS) approximation \cite{Helgaker}.
The dimension of CAS is determined by the number of electrons in $active$ orbitals only.
The notation CAS$(m,n_A)$ indicates filling $m$ electrons in $n_A$ active orbitals.
For our example, six determinantal states built from CAS(2,4) (2 electrons occupying 4 active orbitals)
are shown in Fig.~\ref{fig:StateLabel} (b).
The CAS is designed to unbiasedly capture contributions from determinantal states
made of inactive and active orbitals, which serve as the multi-reference states.

One can further expand the CI space by including states of p-h pairs with respect to multi-reference states defined by CAS.
Following the notation in Ref. \cite{PhysRevB.86.165128}, the expanded CI spaces are denoted by
CAS$(m,n_A)$CIS [single p-h pair upon CAS$(m,n_A)$], CAS$(m,n_A)$CISD [single and double p-h pairs upon CAS$(m,n_A)$],.. etc. 
The generalization is equivalent to the ``restricted active space'' (RAS) approximation \cite{Helgaker}.
In RAS, the inactive orbitals are mostly occupied but are allowed to have small number of holes;
the secondary orbitals are mostly empty but are allowed to have a small number of particles; the active orbitals have no constraints.
We use the notation RAS$(n_I,-k;n_S,l)$ to indicate allowing maximum $k$ holes (we use minus sign to indicate 
the holes) in $n_I$ inactive orbitals, 
and maximum $l$ particles in $n_S$ secondary orbitals. Typically RAS($n_I$,-1;$n_S$,1) = CAS$(m,n_A)$CIS, with
$n_I$ and $n_S$ are respectively the number of inactive and secondary orbitals.

Note that in CAS$(m,n_A)$, we specify the the number of active orbitals
whereas in RAS$(n_I,-k;n_S,l)$ we specify inactive and secondary orbitals, and $n_I+n_S+n_A = n_o$.
The single-reference expansion means no active orbitals.
In Table I we summarize CI truncation schemes based on electron distributions over different orbital classes.
\begin{center}
TABLE I: Electron distributions over different orbital classes for CI truncation schemes.
\begin{tabular}{l|| l| l| l }  \hline
 Scheme &  inactive & secondary  & active  \\  \hline \hline
 CIS(D,T,Q) &  maximum 1 (2,3,4) hole & maximum 1 (2,3,4) electron &  -- \\ \hline
 CAS($m, n_A$) &  0 hole & 0 electron & no constraint \\ \hline
 RAS$(n_I,-k;n_S,l)$ & maximum $k$ hole & maximum $l$ electron & no constraint \\ \hline
\end{tabular}
\end{center}
In terms of the variational principle, different truncation schemes correspond to different subspaces,
and one finds the best single-particle orbitals in the restricted space.
%In the following subsections, we give a criterion when the truncated schemes are valid.
%We further demonstrate that when the truncation is valid, one can find the best orbitals simply by iterations.

%%%%%%%%%%%%%%%%%%%%%%%%%%%%%%%%%%%%%%%%%%%%%%%%%%%%%%%%%%%%%%%%%%%%%%%%%%%%%%%%%%%%%%%%%%%%%%

\subsection{Natural orbitals, degree of correlation, and iteration to ground state}
If the many-body ground state $|GS \rangle$ is known, one can determine the ``natural orbital'' basis
which leads to the most rapid convergence of the CI calculation. Natural orbitals are eigen basis of
the one-particle density matrix 
\beq
\mathbf{D}[\{d\}] =  \langle GS |  d_i^{\dagger} d_j |GS \rangle,
\eeq
with  operators $\{d\}$ specifying the orbitals and $i,j=1,2,...n_o$. The eigenvalues of $\mathbf{D}$ give the natural occupancies.
When only one determinantal state is needed (non-interacting Hamiltonian), 
the natural occupancies are either 1 or 0. Any natural occupancy deviating from 0 or 1
indicates the need of more than one determinantal state to properly describe the ground state.

The degree of correlation can be thus characterized by the minimum number of determinantal states needed
to describe the ground state, which is directly related to the natural occupancies.
Similar to the entanglement entropy \cite{Peschel_11,PhysRevLett.108.087004}, we can define the ``correlation entropy'' as
\beq
S = \sum_{i=1}^{n_o} S_i = -\sum_{i=1}^{n_o} p_i \log p_i,
\label{eqn:S_corr}
\eeq
to measure the degree of correlation. Here $p_i$ is the occupation of $i$th {\em natural orbital}.
This definition ensures that when $p_i = 0$ or 1, the resulting entropy is zero.
Therefore the larger the correlation entropy, the more determinantal states are needed, and
the more correlated the ground state is. 

Since the ground state is not known until the calculation is performed, in practice  the natural orbitals 
are determined in the following way.
Starting from an orbital basis $\{d^{(0)}\}$ (can be arbitrary), the ground state is determined under some CI truncation scheme, from which
the single-particle density matrix ($\mathbf{D}^{(0)}$) and the natural orbitals ($\{d^{(1)}\}$ which diagonalize $\mathbf{D}^{(0)}$) are determined. 
This procedure can be iterated as follows:
\beq
\begin{split}
 \{d^{(0)}\} &\Rightarrow |GS^{(0)} \rangle \Rightarrow 
 \mathbf{D}^{(0)} =  \langle GS^{(0)} |  (d^{(0)}_i)^{\dagger} d^{(0)}_j  |GS^{(0)} \rangle, \,\, d^{(1)}_i =\sum_j  U^{(1)}_{i,j} d^{(0)}_j \\
 \{d^{(1)}\} &\Rightarrow |GS^{(1)} \rangle \Rightarrow 
 \mathbf{D}^{(1)} =  \langle GS^{(1)} |  (d^{(1)}_i)^{\dagger} d^{(1)}_j  |GS^{(1)} \rangle, \,\, d^{(2)}_i =\sum_j  U^{(2)}_{i,j} d^{(1)}_j \\
 .... & .... \\  \{d^{(p)}\} &\Rightarrow |GS^{(p)} \rangle \Rightarrow 
 \mathbf{D}^{(p)} =  \langle GS^{(p)} |  (d^{(p)}_i)^{\dagger} d^{(p)}_j  |GS^{(p)} \rangle,
 \end{split}
 \label{eqn:iteration}
\eeq
where the $\{d^{(i+1)}\}$ ($d^{(i+1)}_i =\sum_j  U^{(i+1)}_{i,j} d^{(i)}_j $) orbitals diagonalize $\mathbf{D}^{(i)}$. 
Some key steps of solving the problem during this successive orbital transformation is given in Appendix A.

%As for application, 
Eq.~\eqref{eqn:iteration} is useful in computing the ground state based on the following observations.
First, we observe that the iteration converges (see Section III B and D), and
the converged ground state corresponds to a variational solution (local energy minimum), subject to the given truncation scheme.
This result can be understood as follows. The natural orbital basis is the single-particle basis which
requires the least number of determinantal states to describe the full $|GS\rangle$, therefore the CI method performs best
in this basis. %a better approximation. % (note that CI truncations discard determinantal states). 
Indeed, we find that during the iterations defined by Eq.~\eqref{eqn:iteration}, more and more natural orbital occupations approach 0 or 1 
(justifying the use of the secondary and inactive orbitals), and the ground state energy is lowered (see Table III and V).

The second observation is that when the CI schemes allow all orbitals to mix during iterations, the 
converged solution is the global minimum subject to the CI scheme.
This is an empirical statement and numerical evidence will be provided in the next section.
This observation implies that a CAS scheme is not suitable for finding the global minimum because
during iterations, secondary, active, and inactive orbitals do not mix \cite{Helgaker}.
%, and the iteration converges to the local minimum of intermixing active orbitals only.
The CIS(D,T) and RAS schemes, however, have no problem finding the global minimum. 
Therefore, {\em if} the ground state falls into the subspace defined
by CIS(D,T) or RAS schemes, one can find the full ground state in the restricted subspace \cite{Grev_1992}.

\subsection{Limitations of the CI truncation}

We conclude this section by discussing the limitations of the CI truncation. 
When none of  the natural occupations approach 0 or 1 (within $10^{-4}$), the CI method does not 
have any meaningful advantage, because the exact ground state involves too many
determinantal states, no matter which orbital basis one uses. 
This limit intrinsically originates from the strong correlation
of the ground state (or the Hamiltonian), where an expansion based on the determinantal basis
is not a good starting point. Examples will be provided in the next section.

\section{Application to a finite-size model}

\subsection{Overview}
In this section we provide two examples demonstrating the efficiency of the CI method.
The efficiency, for our purpose, is quantified by the number of states needed during the calculation.
%, which intrinsically depends on the number of determinantal states needed to capture the full ground state.
The first example is the Anderson impurity model, where the on-site repulsion is present only
at one (impurity) site. In this case, the CI method performs well -- including only a tiny fraction of the total HS 
gives the ground state with high accuracy. The second example is a one-dimensional Hubbard model where
on-site repulsion is present at all sites. In this case the CI method performs poorly -- essentially
all states in the Hilbert space are needed to get the full ground state. 

We consider 8 electrons and 8 spatial orbitals (16 spin-orbitals) where the full ED calculation can be performed with ease. 
The CI schemes are formulated in spin-orbitals, i.e. same spatial orbitals of opposite spins have different labels when 
specifying inactive ($n_I$), active ($n_A$), and secondary ($n_S$) orbitals. In the current case $n_A + n_I + n_S = n_o = 16$. 
Using spin-orbitals facilitates describing the superconducting state, which does not conserve
the total number of particles (see Appendix B). 
The ground state is searched in the $M_z=0$ ($M_z$ is the $z$-component of total spin) 
invariant subspace of dimension 4900 [$(C_4^8)^2$], and is obtained
using Davidson algorithm \cite{Davidson197587}.

\subsection{Anderson impurity model}
The Hamiltonian of the Anderson impurity model is
\beq
H = \varepsilon_d  \sum_{\sigma} c^{\dagger}_{1,\sigma} c_{1,\sigma} + U n_{1, \uparrow} n_{1, \downarrow} 
+ \sum_{p=2}^{8} \sum_{\sigma} t_p [ c^{\dagger}_{1,\sigma} c_{p,\sigma} + h.c.]
+ \sum_{p=2}^{8} \sum_{\sigma} \varepsilon_p  c^{\dagger}_{p,\sigma} c_{p,\sigma} 
\label{eqn:And}
\eeq
We fix the parameters $\varepsilon_p = \pm 2, \pm 4/3, \pm 2/3, 0$ and $t_p = 1/\sqrt{8}$, and
vary $U$ with $\varepsilon_d=-U/2$ (half-filling).
Table II gives the natural occupancies for $U=20, 10, 5 ,2 ,1$. 
The degenerate occupancies are due to spin symmetry. 
\begin{center}
TABLE II: Anderson impurity model. Natural orbital occupancies for $U=20, 10, 5, 2, 1$ and $\mu=U/2$. \\
Due to spin degeneracy, $n_{2i-1} = n_{2i}$ ($i=1,2,...,8$).
\begin{tabular}{l| l| l| l| l|l |l |l |l}  \hline
 $U$ &   $n_1$, $n_2$ & $n_3$, $n_4$ & $n_5$, $n_6$ &  $n_7$, $n_8$ &  $n_9$, $n_{10}$ 
 &  $n_{11}$, $n_{12}$ &  $n_{13}$, $n_{14}$ &  $n_{15}$, $n_{16}$  \\  \hline \hline
 20 &  1 & 0.999999 & 0.999566& 0.582742& 0.417258& 0.000434141& 8.28594e-07& 3.08453e-10 \\  \hline
 10 & 1& 0.999999& 0.999197& 0.67482& 0.32518& 0.000802745& 1.45276e-06& 1.03073e-09 \\ \hline
 5 & 1& 0.999999& 0.999331& 0.81756& 0.18244& 0.000669318& 1.25351e-06& 1.06296e-09 \\ \hline
 2 & 1& 1& 0.999826& 0.954373& 0.0456267& 0.000174072& 3.46593e-07& 2.57253e-10 \\ \hline
 1 &1& 1& 0.999954& 0.987583& 0.0124174& 4.64471e-05& 9.20178e-08& 8.14356e-11 \\ \hline
\end{tabular}
\end{center}
From Table II we see that for a wide range of $U$ values, out of 16 spin-orbitals, only eight orbitals -- from 5 to 12  
-- are active (partially occupied), and one in principle only needs determinantal states composed of 8 orbitals
to accurately describe the ground state. %From Table II, t
The intermediate $U = 10$ needs the largest number of determinantal states, 
we therefore apply several CI schemes for $U=10$ to Eq.~\eqref{eqn:iteration}. 
The CI schemes, their dimensions, and the ground state energy at each iteration are given in Table III.
The starting orbital basis is chosen as the local basis which is very different from the natural basis. 
We have tried several starting orbital basis, and they all converge to the same ground state energy.
\begin{center}
TABLE III: Calculated ground state energy using iterated orbital basis for different CI schemes.\\
These results are for $U=10$ and $\varepsilon_d=-5$. The number in parenthesis indicates the dimension \\of the CI subspace.
``-'' indicates the same energy as the previous iteration.
\begin{tabular}{l| l| l |l | l | l}  \hline
iteration &  CISD (361)  & CISDT (1545) &  RAS(4,-1; 4,1) (523)  & RAS(6,-1; 6,1) (118) & Exact (4900)   \\  \hline \hline
0 &  -13.2455 &  -13.2465 & -13.1178   & -13.1167  &-13.2465  \\ \hline
1 &  -13.2464 &  - & -13.2465   &  -13.2463  &- \\ \hline
2 &  - &  - &  -  &  -13.2464  &- \\ \hline
3 &  - &  - &  - &  -  &-  \\ \hline
 \end{tabular}
\end{center}

We can point out some general features. First, the ground state energy decreases during iterations, 
indicating that it converges to at least a local minimum in the subspace. Moreover,
the converged ground state energy is very close to the exact one, indicating
the solution is the {\em global} minimum in the subspace (the statements in Section II.C). 

Depending on the required accuracy, all four CI schemes give good ground state energy with the error smaller than $10^{-4}$.
Typically, RAS schemes perform better than a single-reference CI. Once the natural orbital basis is obtained,
we can perform the calculation again, using the CAS scheme based on the natural orbital occupation.
We found that performing CAS(4,8), which contains only 35 determinantal states ($C^8_4$ plus $M_z=0$ constraint), after
obtaining natural orbitals using $any$ schemes listed in Table III
%CISDT, RAS(4,-1; 4,1), and even RAS(6,-1; 6,1)  
 gives the practically exact ground state energy of -13.2465.
 This means that one in principle only needs to  keep 118 states [RAS(6,-1; 6,1) in Table III] out of
 4900 states during the entire iteration to get the ground state of high accuracy, although in practice 
 one needs to check the convergence against other CI schemes keeping more states.
 Based on the natural occupancy analysis (Table II), 35 is the $minimum$ number of 
determinantal basis states needed.
%Switching to CAS scheme, which further truncates the CI space, benefits the calculation of Green's function.

\subsection{Strategy to compute the ground state and Green's functions}
 The impurity Green's function is defined as
\beq
\begin{split}
G_{imp}(i \omega_n) &= \langle GS| c_1 \left[ i \omega_n - ( H - E_{GS} )    \right]^{-1} c_1^{\dagger} |GS\rangle \\
&+ \langle GS| c_1^{\dagger} \left[ i \omega_n + ( H - E_{GS} )    \right]^{-1} c_1 |GS\rangle.
\end{split}
\eeq
with $H$ defined in Eq.~\eqref{eqn:And} and $E_{GS}$ the ground state energy. 
To compute $G_{imp}$, one needs to specify the HS of $n_e \pm 1$ particles (see also Appendix A).
Expressing the ground state with a small number of determinantal states can accelerate the Green function calculation, 
since the number of corresponding $(n_e \pm 1)$-particle states also reduces.
We now formulate a general and efficient  strategy  to compute the ground state and impurity Green's function.

\bi
\item Use RAS to obtain the natural orbital basis. To save time, one starts by using severely truncated scheme, then
checks the convergence by comparing the ground state energy obtained from the less-truncated CI schemes.

\item Once the natural orbital basis is determined, perform a CAS calculation [according to the natural occupancies]
to further reduce the dimension of the subspace without losing accuracy.
%This step is to use the minimum number of states to describe the ground state  
The reason is that the corresponding $n_e\pm 1$ Fock spaces are also small, accelerating the Green's function calculation.

\item With the ground state in the CAS scheme \cite{RAS}, compute the Green's function using the Lanczos algorithm \cite{Grosso, RevModPhys.66.763}. 
\ei

%%%%%%%%%%%%%%%%%%%%%%%%%%%%%%%%%%%%%
\subsection{Hubbard model on a ring}
We test this strategy by considering a Hubbard model on an 8-site  ring described by the following Hamiltonian:
\beq
H = -\mu  \sum_{i, \sigma} c^{\dagger}_{i,\sigma} c_{i,\sigma} + U \sum_{i, \sigma} n_{i, \uparrow} n_{i, \downarrow} 
-t  \sum_{i=1,\sigma}^{8}   [ c^{\dagger}_{1,\sigma} c_{i+1,\sigma} + h.c.],
\label{eqn:1DH}
\eeq
with the periodic boundary condition $c_9 \equiv c_1$. We take $t=1$ for different values of $U$ at half filling ($\mu=U/2$).
In Table IV we list the natural occupancies for $U=20,10,4,2$ from the full ED calculation.
Note that the inversion symmetry is responsible for the degeneracy in addition to spin.
\begin{center}
TABLE IV: 8-site ring Hubbard model. Natural orbital occupancies for $U=20,10,4,2$ and $\mu=U/2$. \\
Due to spin degeneracy, $n_{2i-1} = n_{2i}$ ($i=1,2,...,8$).
\begin{tabular}{l| l| l| l| l|l |l |l |l}  \hline
 $U$ &   $n_1$, $n_2$ & $n_3$, $n_4$ & $n_5$, $n_6$ &  $n_7$, $n_8$ &  $n_9$, $n_{10}$ 
 &  $n_{11}$, $n_{12}$ &  $n_{13}$, $n_{14}$ &  $n_{15}$, $n_{16}$  \\  \hline \hline
 20 &0.636333& 0.59957& 0.59957& 0.5& 0.5& 0.40043& 0.40043& 0.363667  \\  \hline 
 10 & 0.748239& 0.695164& 0.695164& 0.5& 0.5& 0.304836& 0.304836& 0.251761 \\ \hline
 4 & 0.914041& 0.882618& 0.882616& 0.500001& 0.499999& 0.117384& 0.117382& 0.0859591\\ \hline
 2 & 0.974145& 0.963764& 0.963764& 0.500002& 0.499998& 0.0362365& 0.0362365& 0.0258546 \\ \hline
\end{tabular}
\end{center}
From Table IV, we see that the natural occupancies are not close to either 0 or 1, even for $U=2$.
From the discussion in Section II.C and D, we conclude that the CI method is $not$ suitable for this situation. 
For $U=2$ case, we perform CISDTQ (=CI[4 p-h]), CI[5 p-h], CI[6 p-h] calculations (the number in square brackets indicates
the maximum number of p-h pairs), and find that the efficiency of the CI method is poor -- to converge to $10^{-4}$ in energy,
the required number of determinantal states is of the order of the original HS. 
The results are summarized in Table V. One notices that the iteration still finds the global minimum
in the restricted space.

\begin{center}
TABLE V: Calculated ground state energy using iterated orbital basis for different CI schemes.\\
These results are for $U=2$ and $\mu=U/2$. The number in parenthesis indicates the dimension \\of the CI space.
``-'' indicates the same energy as the previous iteration.
\begin{tabular}{l| l| l |l | l }  \hline
iteration &  CISDTQ (3355) & CI[5 p-h] (4539)  & CI[6 p-h] (4867)   & Exact (4900)   \\  \hline \hline
0 &  -14.1328 & -14.5148 &-14.5681   &-14.5682  \\ \hline
1 &  -14.5581 & -14.5642 &-14.5682   &- \\ \hline
2 &  -14.5625 & - & -  &- \\ \hline
3 &  - &  - &  - &  -   \\ \hline
 \end{tabular}
\end{center}

%This example shows that the use of CI method is limited by the strong correlation effect -- if
%the ground state involves too many determinantal states, the CI method is not a good choice. 
As was mentioned in the introduction, the DMFT maps a lattice problem onto an impurity one. 
By comparing two examples in Section III, the lattice model is intrinsically more complicated
because its ground state involves many more determinantal states. This provides a quantitative explanation why the impurity 
model is more numerically tractable than the lattice model. 
Within DMFT, our two model calculations suggest that the main advantage of CI method over the full ED is the ability to include more
bath orbitals.

%%%%%%%%%%%%%%%%%%%%%%%%%%%%%%%%%%%%%%%%%%%%%%%%%%%%%%%%%%%%%%%%
\section{Application to dynamical mean field theory}

\subsection{Overview}
We now apply the CI method introduced in Section II to the DMFT, using ED as the impurity solver.
The DMFT involves self-consistently determining the parameters of an impurity problem,
and CI method enters as an approximation to the ED solver. 
%T is computing
The impurity Green's function, which is the most resource-consuming part of DMFT equations (in terms of time and computer memory),
is computed using the strategy formulated in Section III C.
We choose to solve the Hubbard model with attraction on the infinite-dimension Bathe lattice (bandwidth $4t$), 
under the s-wave superconducting self-consistent condition.
This model has been studied in Refs \cite{PhysRevLett.86.4612, PhysRevLett.88.126403, Toschi_05}.
The main complication arising from the superconducting state is that the total particle number is not conserved anymore,
and one has to formulate the problem in the grand-canonical ensemble including particle number fluctuations.
The grand-canonical ensemble enlarges the HS dimension and complicates the calculation, which we choose as a platform to test the
efficiency of the CI method.

We solve the problem at half and quarter fillings.  
%Away from half-filling, an extra self-consistency loop over the chemical potential is needed. 
%Using the strategy stated in Section III. C, W
We are able to consider up to 9 bath sites (totally 10 sites including the impurity)
using moderate computational resources (a single-core laptop with 2GB memory).
The convergence can be accelerated by using severely truncated CI schemes in the first few DMFT self-consistency iterations,
as these iterations improve the bath parameters at a relatively low computational cost.
We found that the results for 5 bath sites are essentially identical
to those with 7 and 9 bath sites, with the exception of the spectral functions \cite{Toschi_05}. 
For spectral functions, however, the gap (or more precisely the reduction of the density of states) 
around the Fermi energy is robust for 5, 7, and 9 bath sites.
As for the ED solver, 
we use the fictitious temperature of 0.1$t$, and 
Levenberg-Marquardt algorithm \cite{Minpack} to determine the bath parameters.
In the next subsection we will formulate the self-consistent equation and show the results.
Details of solving the impurity model in the grand-canonical ensemble are given in the Appendix B.

\subsection{Self-consistent equation}

The self-consistent equation for the s-wave superconducting state has been derived in Refs \cite{PhysRevLett.72.1545, RevModPhys.68.13}.
To describe the superconducting state, one uses the Nambu Green's function 
which breaks the U(1) symmetry. By defining a Nambu spinor as
$\Psi^{\dagger}_{\mathbf{k}} = (c^{\dagger}_{\mathbf{k}, \uparrow}, c_{-\mathbf{k}, \downarrow})$, 
the lattice Green function (a 2$\times$2 matrix) in imaginary time is given by
\beq
\begin{split}
\hat{G} (\mathbf{k}, \tau) &= -T \langle  \Psi_{\mathbf{k}}(\tau) \Psi^{\dagger}_{\mathbf{k}}(0) \rangle=
-T \langle \begin{pmatrix} c_{\mathbf{k}, \uparrow}(\tau) \\ c^{\dagger}_{-\mathbf{k}, \downarrow}(\tau)   \end{pmatrix} 
\begin{pmatrix} c^{\dagger}_{\mathbf{k}, \uparrow} (0) & c_{-\mathbf{k}, \downarrow}(0) \end{pmatrix}  
  \rangle  \\
&= \begin{pmatrix} -T \langle c_{\mathbf{k}, \uparrow}(\tau) c^{\dagger}_{\mathbf{k}, \uparrow}(0) \rangle
&  -T \langle c_{\mathbf{k}, \uparrow}(\tau) c_{-\mathbf{k}, \downarrow}(0) \rangle \\
-T \langle c^{\dagger}_{-\mathbf{k}, \downarrow}(\tau) c^{\dagger}_{\mathbf{k}, \uparrow}(0) \rangle&
 -T \langle c^{\dagger}_{-\mathbf{k}, \downarrow}(\tau) c_{-\mathbf{k}, \downarrow}(0) \rangle
  \end{pmatrix}  \\
  & \equiv
  \begin{pmatrix} G(\mathbf{k}, \tau) & F(\mathbf{k}, \tau) \\ F(-\mathbf{k}, -\tau)^* & -G(-\mathbf{k},-\tau) \end{pmatrix}
  = \begin{pmatrix} G(\mathbf{k}, \tau) & F(\mathbf{k}, \tau) \\ F(\mathbf{k}, \tau)^* & -G(-\mathbf{k},-\tau) \end{pmatrix}
  \end{split}
  \label{eqn:Nambu_SC}
\eeq
where $\langle ... \rangle$ represents the ground state expectation value. 
The matrix structure of $\hat{G} (\mathbf{k}, \tau)$ is obtained by using time-translational symmetry
and anti-communication relation.
%For relation between two diagonal terms, we have used (neglecting spin and momentum subscript) 
%\begin{equation*}
%\begin{split}
%& \langle c(\tau) c^{\dagger}(0) \rangle = g(\tau) \\ &\Rightarrow
%\langle c^{\dagger}(\tau) c(0) \rangle =  -\langle c(0) c^{\dagger}(\tau)  \rangle
%= -\langle c(-\tau) c^{\dagger}(0)  \rangle = -g(-\tau).
%\end{split}
%\end{equation*}
For example, two off-diagonal terms are related (neglecting the spin and momentum subscripts):
\begin{equation*}
\begin{split}
& \langle c(\tau) c(0) \rangle \equiv f(\tau) \\ &\Rightarrow
\langle c^{\dagger}(\tau) c^{\dagger}(0) \rangle =  \langle \left[ c(0) c(\tau) \right]^{\dagger} \rangle
= \langle \left[ c(-\tau) c(0) \right]^{\dagger} \rangle = f^*(-\tau).
\end{split}
\end{equation*}
For the third line in Eq.~\eqref{eqn:Nambu_SC}, we further use the property that
for a singlet pairing field, $F(-\mathbf{k}, -\tau) =F(\mathbf{k}, \tau) $ [for triplet pairing (p-wave),
one instead obtains $F(-\mathbf{k}, -\tau) =-F(\mathbf{k}, \tau)$] \cite{RevModPhys.68.13}. 
Under the single-site DMFT approximation, where self energy has no momentum dependence, the lattice Green's function
at Matsubara frequencies is
\beq
\hat{G}( \mathbf{k}, i \omega_n)^{-1} = 
\begin{pmatrix} i \omega_n - (\varepsilon_{\mathbf{k}} - \mu)  & 0 \\ 
0 &  i \omega_n + (\varepsilon_{-\mathbf{k}} - \mu)  \end{pmatrix}
- \hat{\Sigma} (i \omega_n),
\eeq
with $\hat{\Sigma}( i \omega_n) =
\begin{pmatrix} \Sigma(i \omega_n) & S( i \omega_n) \\ S(i \omega_n) & -\Sigma(i \omega_n)^* \end{pmatrix}$ 
computed from the impurity problem.
The local Green's function
\beq
\hat{G}_{loc} ( i \omega_n)=
\frac{1}{N}\sum_{\mathbf{k}} \hat{G}( \mathbf{k}, i \omega_n) = 
\begin{pmatrix} G(i \omega_n) & F( i \omega_n) \\ F(i \omega_n) & -G(-i \omega_n)  \end{pmatrix},
\eeq
is required to be the impurity Green's function by the self-consistent condition. 

\begin{figure}[http]
\epsfig{file=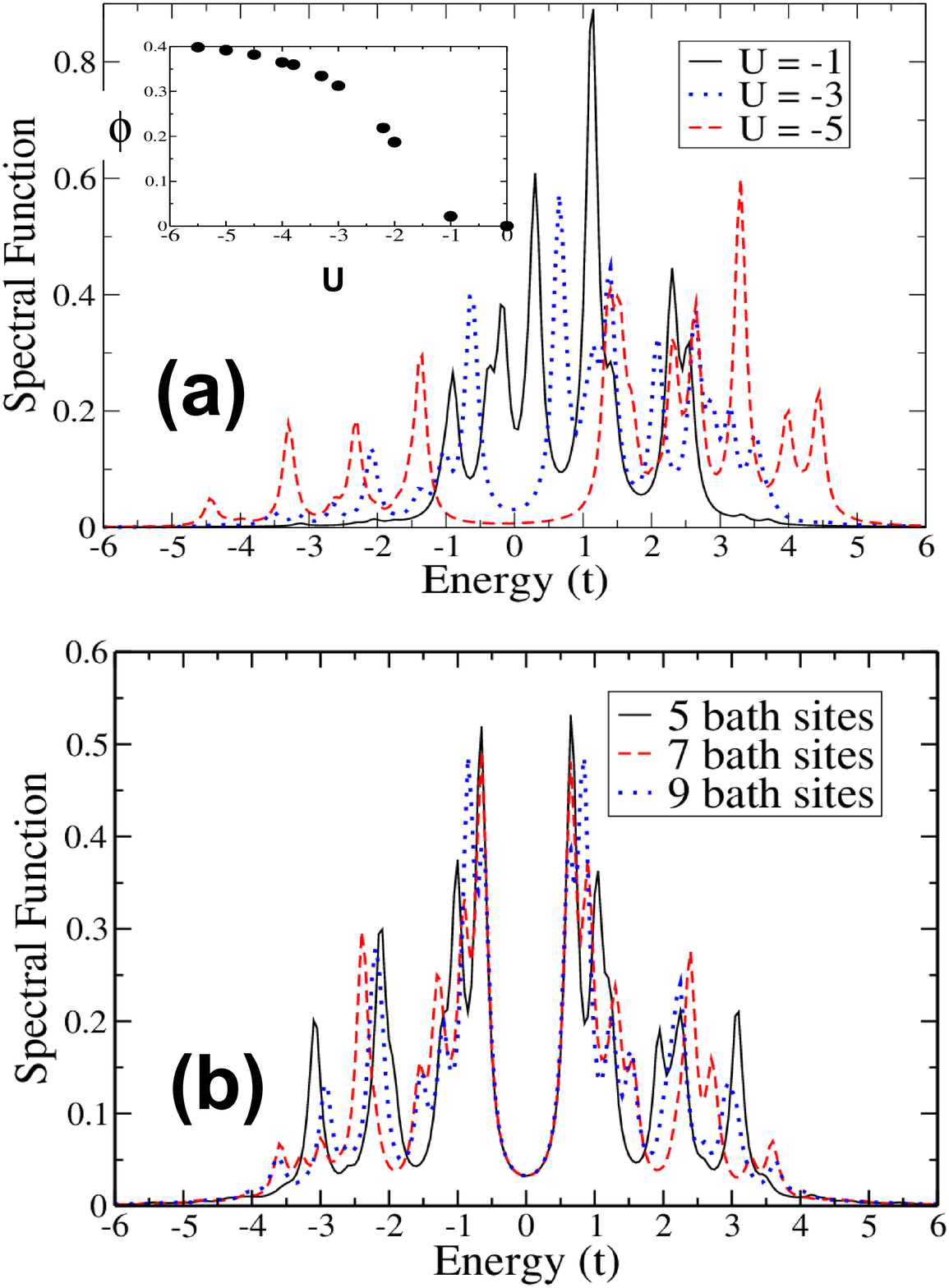, width = 0.4\textwidth}
\caption{(Color online) (a) The spectral function of the spin-up electron for $U=-1, -3, -5$ (5 bath sites)
at quarter filling. A gap at the Fermi energy  opens when $|U|$ increases. (Inset) The corresponding order parameter $\phi$
as a function of $U$. (b)  The spectral function of the spin-up electron for $U=-3$ for 5, 7, 9 bath sites.
The gap structures around Fermi energy are almost identical.
}
\label{fig:SuperCond}
\end{figure}

The pairing order parameter is defined as $\phi = T \sum_n F( i \omega_n)$.
For any negative $U$, $|\phi|$ is non-zero \cite{Toschi_05}. When $|U|$ is small, the order parameter
is exponentially small and cannot be resolved in our calculation. 
The calculated spectral function at quarter filling, using 5 bath sites for $U=-1,-3,-5$ is shown in Fig.~\ref{fig:SuperCond} (a).
We see that at the Fermi energy, there is a reduction in the density of states accompanied by two sharp peaks on both
sides. The reduction in the spectral function increases (eventually developing a gap) as $|U|$ becomes larger. 
This corresponds to the formation of the superconducting gap and its quasi-particle excitations.
Using the ED solver, however, we cannot observe the gap of too small amplitudes due to the finite bath discretization. 
In the inset we show the pairing order parameter as a function of $U$.
Fig.~\ref{fig:SuperCond} (b) shows the spectral function at half-filling for $U=-3$ using 5, 7, and 9 bath sites.
%We do not impose particle-hole symmetry in the self-consistency loop.
We find that the only change upon increasing the number of the bath sites is the spectral function. 
The quantities at the imaginary frequencies, as well as the order parameter, are well converged (the relative change 
$|\Delta \phi|/|\phi|$ in the order parameters
is smaller than $10^{-4}$).
From Fig.~\ref{fig:SuperCond} (b), we see that the spectral functions around the Fermi level, 
especially the gap structure, do not change as the bath size increases.  
Finally, we mention (not shown) that the same self-consistent equations can apply to the positive $U$ regime.
There the pairing order parameter vanishes, and known results such as the metal-insulator transition are recovered.

\section{Conclusions}

In this paper, we analyze the efficiency of the configuration interaction method as an approximation to
the full exact diagonalization. Not surprisingly, the CI method performs well if the exact ground state can be
described by a small number of determinantal states; but does not hold any meaningful advantage (compared to full ED)
when the ground state is composed of many determinantal states. In terms of the variational principle, the validity of the CI truncation
depends intrinsically on the overlap between the exact ground state and the CI truncated subspace.
The performance can be estimated by examining the natural occupancies 
computed by a truncated CI scheme (not the full ED). If there are natural orbitals with occupations close to 1 and 0,  
the CI method is useful, otherwise it is not. We illustrate this point
with two examples, the Anderson impurity model and the Hubbard model.
When the CI truncation is valid, we demonstrate that by iteratively searching for the natural orbital basis under the CI schemes
that allow all orbitals to mix (such as CISD or RAS, but $not$ CAS), a very accurate ground state can be obtained.
This procedure can be viewed as searching for the ground state in the variational subspace.
%whose dimension can be just 10$\%$ (even smaller) of the full Hamiltonian. 
Once the natural orbitals are determined, a CAS can be performed to express the ground state in the minimum number
of determinantal states without losing accuracy. The impurity Green's function can then be efficiently
computed from the CAS ground state.

As an illustration, we apply this method to the single-site dynamical mean field theory using ED as the impurity solver. 
We solve the Hubbard model with attraction in the grand-canonical ensemble ($i.e.$ including the particle number
fluctuation) for the s-wave superconducting state. All expected results, such as non-zero pairing and
gap opening, are reproduced. Moreover, we are able to check the convergence with respect to the bath size
(up to 9 bath sites) using only moderate computational resources (a single-core laptop of 2GB memory). 

Our two model calculations, Anderson impurity and ring Hubbard model, suggest that CI method is intrinsically
efficient for the impurity model (only a few orbitals are correlated), but not so for the lattice model (all orbitals are correlated). 
Based on this observation, we think the CI method is most useful, in terms of studying correlated systems, when 
combining with DMFT.
As a DMFT impurity solver, the main advantage of CI method over the full ED is its ability to include more
bath orbitals. This is consistent with our superconducting calculation and the cluster calculation done in Ref. \cite{PhysRevB.86.165128}.
The CI solver, by itself, is also efficient when the impurity problem is far away from the half filling.

\section*{Acknowledgements}
C.L. thanks Hung The Dang, Ara Go, Andrew Millis, Takashi Oka, Hsiang-Hsuan Hung,
and Andreas R\"uegg for helpful discussions. We thank Richard Hatch, Hosung Seo, and Alex Slepko for insightful comments.
Support for this work was provided through Scientific Discovery through Advanced Computing (SciDAC) program 
funded by U.S. Department of Energy, Office of Science, Advanced Scientific Computing Research and Basic Energy Sciences under award number DESC0008877.

\bibliography{CI_DMFT}

\begin{thebibliography}{41}
\bibitem[{p-h()}]{p-h}
\bibinfo{note}{Note that when mentioning states of non-zero particle-hole
  pairs, a reference state is already (implicitly) defined.}
\bibitem[{RAS()}]{RAS}
\bibinfo{note}{If the CAS space is inadequate because no natural occupancies
  approach close enough 1 and 0, one then uses the RAS space of the smallest
  dimension.}
\end{thebibliography}

\appendix

\section{Successive orbital transformations}
We assume the Hamiltonian is specified in the basis  $\{ d^{(0)} \}$ as $H(\{ d^{(0)} \})$.
Two sets of  basis orbitals, $\{ d^{(0)} \}$  and $\{ d^{(p)} \}$, are related by an unitary  matrix 
\beq
d^{(p)}_i = \sum_j \tilde{U}^{(p)}_{i,j} d^{(0)}_j.
\eeq
Once $\tilde{U}^{(p)}$ is known, one can express $H(\{ d^{(0)} \})$ in terms of $H(\{ d^{(p)} \})$,
and re-diagonalize the problem in the $\{ d^{(p)} \}$ basis.
As shown in Eq.~\eqref{eqn:iteration},  $\tilde{U}^{(p)}$  can be obtained by 
\beq
\begin{split}
\tilde{U}^{(p)} &= U^{(p)}  ...  U^{(2)}U^{(1)} \equiv \Pi_{i=1}^{p} U^{(i)}  \\
&= U^{(p)} \Pi_{i=1}^{p-1} U^{(i)} = U^{(p)}  \tilde{U}^{(p-1)}.
\end{split}
\eeq
Note that in our notation, $U^{(p)}$ connects $\{ d^{(p-1)} \}$  and $\{ d^{(p)} \}$ basis sets, whereas 
$\tilde{U}^{(p)}$ connects $\{ d^{(0)} \}$  and $\{ d^{(p)} \}$ basis sets.
For the first iteration, $\tilde{U}^{(0)}$ can be arbitrary.
At the iteration using $\{ d^{(p)} \}$ basis set [see Eq.~\eqref{eqn:iteration}], $\tilde{U}^{(p-1)}$ is needed.
After getting the CI ground state $|GS^{(p)} \rangle$, we diagonalize
the single particle density matrix $\mathbf{D}^{(p)}$ and 
get $U^{(p)}$, from which $\tilde{U}^{(p)} = U^{(p)}  \tilde{U}^{(p-1)}$ is obtained for the
next iteration using the $\{ d^{(p+1)} \}$ basis set.

%In any model calculation, the form of $H(\{ d^{(0)} \})$ is given, 
%Therefore in the implementation, only $\tilde{U}^{(p)}$ is needed.

We would like to point out an important step concerning the evaluation of the single-particle density matrix 
in the truncated CI space [$\mathbf{D}^{(p)}$ in Eq.~\eqref{eqn:iteration}].
Since the CI ground state $| GS^{(p)} \rangle$ does not include all $n_e$-particle Hilbert space, 
the state $ (d^{(p)}_i)^{\dagger} d^{(p)}_j |GS^{(p)} \rangle$
generally leads to components outside the truncated CI space. This numerical error can 
lead to a non-Hermitian $\mathbf{D}^{(p)}$ and therefore non-unitary  $U^{(i)}$.
To avoid this error, we construct the $(n_e-1)$-particle CI space which contains an extra hole of the given $n_e$-particle CI space.
For example, if the $n_e$-particle CI ground state is approximated in RAS$(n_I,-k; n_S, l)$, then 
we construct RAS$(n_I,-k-1; n_S, l)$, which contains $all$ components of
$ d^{(p)}_j |GS^{(p)} \rangle$. The $\mathbf{D}^{(p)}$ is obtained by computing the inner-product of 
$ d^{(p)}_i |GS^{(p)} \rangle$ and $ d^{(p)}_j |GS^{(p)} \rangle$.

In the Green's function calculation, we need to construct both 
RAS$(n_I,-k-1; n_S, l)$ [an $(n_e-1)$-particle space containing all $d^{(p)}_i |GS^{(p)} \rangle$ components] and 
RAS$(n_I,-k; n_S, l+1)$ [an $(n_e+1)$-particle space containing all $(d^{(p)}_i)^{\dagger} |GS^{(p)} \rangle$ components],
to completely describe hole and particle excitations. 
Once they are constructed, the corresponding hole and particle  Krylov spaces respectively spanned by
\beq
\begin{split}
\{ |\psi'_0 \rangle &= c_1 |GS^{(p)} \rangle, |\psi'_1 \rangle = H |\psi'_0 \rangle, ..., |\psi'_{i+1} \rangle = H |\psi'_i \rangle,... \}, \\
\{ |\psi_0 \rangle &= c^{\dagger}_1 |GS^{(p)} \rangle, |\psi_1 \rangle = H |\psi_0 \rangle, ..., |\psi_{i+1} \rangle = H |\psi_i \rangle,... \}
\end{split}
\eeq
are iteratively generated within the $(n_e \pm 1)$ subspace, in which the Green's function is 
computed using the Lanczos algorithm.
Certainly the smaller $n_e$-particle CI space leads to the smaller corresponding $(n_e \pm 1)$-particle CI space and thus
faster calculations. Reducing the dimension of the CI ground state can accelerate the Green's function calculation.
This advantage is magnified in the DMFT calculation.

\section{Grand canonical ensemble and Bogoliubov particles}
Here we formulate the impurity model using the grand canonical ensemble that
can describe the superconducting state.
To describe the superconducting state, the impurity model is generalized to [compare to Eq.~\eqref{eqn:And}]
\beq
H_{And,SC} = \varepsilon_d  \sum_{\sigma} c^{\dagger}_{1,\sigma} c_{1,\sigma}
+ \sum_{p=1}^N t_{sc,p} (c^{\dagger}_{p,\uparrow} c^{\dagger}_{p,\downarrow} + h.c.)
+ U n_{1, \uparrow} n_{1, \downarrow} 
+ \sum_{p=2, \sigma}^{N}  t_p [ c^{\dagger}_{1,\sigma} c_{p,\sigma} + h.c.]
+ \sum_{p=2, \sigma}^{N}  \varepsilon_p  c^{\dagger}_{p,\sigma} c_{p,\sigma},
\label{eqn:And_SC}
\eeq
which breaks the U(1) symmetry via the term $t_{sc,p} (c^{\dagger}_{p,\uparrow} c^{\dagger}_{p,\downarrow} + h.c.)$.
We have assumed site 1 to be the impurity site, and $N$ to be the total number of sites.
Using the standard procedure \cite{Allen_SC}, one performs the particle-hole transformation on one of the spin species and
rewrites Eq.~\eqref{eqn:And_SC} as
\beq
\begin{split}
H_{And,SC} &= \varepsilon_d  ( c^{\dagger}_{1,\uparrow} c_{1,\uparrow} - c_{1,\downarrow} c^{\dagger}_{1,\downarrow})
+ \sum_{p=1}^N t_{sc,p} (c^{\dagger}_{p,\uparrow} c^{\dagger}_{p,\downarrow} + h.c.)
+ U c^{\dagger}_{1,\uparrow} c_{1,\uparrow} c^{\dagger}_{1,\downarrow} c_{1,\downarrow} \\
&+ \sum_{p=2}^{N} t_p  [ c^{\dagger}_{1,\uparrow} c_{p,\uparrow} - c_{1,\downarrow} c^{\dagger}_{p,\downarrow} + h.c.]
+ \sum_{p=2}^{N} \varepsilon_p  (c^{\dagger}_{p,\uparrow} c_{p,\uparrow} - c_{p,\downarrow}  c^{\dagger}_{p,\downarrow} )+ K,
\end{split}
\label{eqn:And_SC2}
\eeq
with the constant $K = \sum_{p=1}^{N} \varepsilon_p$. When computing the Green's function,
the constant $K$ does not play any role.
Using a Nambu spinor, the quadratic part of $H_{And,SC}$ can be expressed as
\beq
\begin{pmatrix} c^{\dagger}_{1,\uparrow} & c_{1,\downarrow}& ... & c^{\dagger}_{N,\uparrow}&   c_{N,\downarrow} \end{pmatrix} H_{quad} 
\begin{pmatrix}
   c_{1,\uparrow} \\  c^{\dagger}_{1,\downarrow}\\ \vdots\\ c_{N,\uparrow} \\ c^{\dagger}_{N,\downarrow}  \end{pmatrix}
   =  \begin{pmatrix} \gamma^{\dagger}_{1} & ... & \gamma^{\dagger}_{2N} \end{pmatrix} U^{T} H_{quad} U \begin{pmatrix}
   \gamma_{1}\\ \gamma_{2} \\ \vdots\\ \gamma_{2N}  \end{pmatrix}.
\eeq
Here $U$ is a $2N \times 2N$ unitary matrix, and one can choose it freely to simplify the calculation.
Note that $\gamma^{\dagger}_i$, which is a linear combination of  $c^{\dagger}_{\uparrow}$ and $c_{\downarrow}$, 
obeys the same anti-commutation relation of fermions, i.e. 
$\{ \gamma_i^{\dagger}, \gamma_j \} = \delta_{ij}$ and $\{ \gamma_i, \gamma_j \} = 0$, and can be
used to specify determinantal basis states.
%Using CI solver, one needs to express the interaction in terms of $\gamma$ operators.
%In order to do this, we need
%$c^{\dagger}_{1,\uparrow} = \sum_{j=1}^{2N} U_{1j} \gamma_j^{\dagger}$ (therefore $c_{1,\uparrow} = \sum_{j=1}^{2N} U_{1j} \gamma_j$),
%and $c_{1,\downarrow} = \sum_{j=1}^{2N} U_{2j} \gamma_j^{\dagger}$ (therefore $c^{\dagger}_{1,\downarrow} = \sum_{j=1}^{2N} U_{2j} \gamma_j$).
%Note that  $c_{1,\uparrow}, c^{\dagger}_{1,\downarrow}$ are expressed solely in $\gamma$, whereas
% $c^{\dagger}_{1,\uparrow}, c_{1,\downarrow}$ are expressed solely in $\gamma^{\dagger}$.
%\subsubsection{Specifying the Hilbert space}
%Because $\{ \gamma_i \}$ satisfies all fermionic anti-commutation relation, one can specify
%the determinantal basis states using $\{ \gamma_i \}$ operators.
We refer to $\gamma_i^{\dagger}$ as a creation operator of a Bogoliubov particle. 
Note that the interaction conserves the number Bogoliubov particles, and
we consider the Hilbert space of $N$ Bogoliubov particles, which is the largest
invariant subspace of $2N$ orbitals. 

As an example, we take $N=4$ (3 bath sites) which generates 8 ($2\times 4$, 2 for spin) Bogoliubov orbitals.
The space of 4 Bogoliubov particles has the dimension of $C^8_4=70$, and the basis state is expressed as
$ \gamma_i^{\dagger} \gamma_j^{\dagger} \gamma_k^{\dagger} \gamma_l^{\dagger} |0\rangle$.
%Now comment the vacuum state $|0\rangle$ and the relation between
%space composed of Bogoliubov particles and original ones. 
Note the vacuum state is a state of zero $Bogoliubov$ particles, not a state of zero ``real'' particles. 
For a given impurity Hamiltonian, one actually
performs the particle-hole transformation on spin down electrons. In terms of original particles,
it implies the proper vacuum state fills all spin-down electrons, i.e.
\beq
|0\rangle = \Pi_{i=1}^{N(=4)} c^{\dagger}_{i,\downarrow} | \text{no particle} \rangle
= c^{\dagger}_{1,\downarrow} c^{\dagger}_{2,\downarrow} c^{\dagger}_{3,\downarrow} c^{\dagger}_{4,\downarrow} | \text{no particle} \rangle.
\eeq
A state in subspace specified by $N (=4)$ Bogoliubov particles is
$\gamma_i^{\dagger} \gamma_j^{\dagger} \gamma_k^{\dagger} \gamma_l^{\dagger} \left[ \Pi_{i=1}^{N(=4)} c^{\dagger}_{i,\downarrow} 
| \text{no particle} \rangle \right]$.
Since $\gamma^{\dagger} \sim c^{\dagger}_{\uparrow} + c_{\downarrow}$, 
when expressing in the original $\{ c_i \}$ basis, it contains subspace of 0, 2, 4, 6, 8 original particles whose 
representatives are
\beq
\begin{split}
 |\text{0 particle} \rangle &= | \text{no particle} \rangle [1] \\
 |\text{2 particle} \rangle &=  c^{\dagger}_{1,\uparrow} c^{\dagger}_{1,\downarrow} | \text{no particle} \rangle [16]\\
 |\text{4 particle} \rangle &=  c^{\dagger}_{1,\uparrow} c^{\dagger}_{2,\uparrow} c^{\dagger}_{1,\downarrow} c^{\dagger}_{2,\downarrow} 
 | \text{no particle} \rangle   [36]\\
 |\text{6 particle} \rangle &= c^{\dagger}_{1,\uparrow} c^{\dagger}_{2,\uparrow} c^{\dagger}_{3,\uparrow} 
   c^{\dagger}_{1,\downarrow} c^{\dagger}_{2,\downarrow} c^{\dagger}_{3,\downarrow}
 | \text{no particle} \rangle [16]\\
 |\text{8 particle} \rangle &= c^{\dagger}_{1,\uparrow} c^{\dagger}_{2,\uparrow} c^{\dagger}_{3,\uparrow}  c^{\dagger}_{4,\uparrow} 
   c^{\dagger}_{1,\downarrow} c^{\dagger}_{2,\downarrow} c^{\dagger}_{3,\downarrow} c^{\dagger}_{4,\downarrow} | \text{no particle} \rangle [1]
\end{split}
\eeq
where the number in the square brackets indicates the dimension. The sum of all dimensions is exactly 70 -- which is the dimension of subspace
of 4 Bogoliubov particles.

\end{document}